\documentclass[prl,twocolumn,showpacs,floatfix,titlepage,superscriptaddress]{revtex4-1}

\usepackage{bm}
\usepackage{graphics}
\usepackage{graphicx}
\usepackage{latexsym}
\usepackage{amsmath}
\usepackage{amsfonts}
\usepackage{amssymb}
\usepackage{amsthm}
\usepackage{dcolumn}
\usepackage{color}
\usepackage{mathrsfs}
\usepackage[left]{lineno}
\usepackage{dcolumn}

\begin{document}

\title{Topological transition of Dirac points in a microwave experiment}

\author{Matthieu Bellec}
\author{Ulrich Kuhl}
\affiliation{Universit\'{e}  Nice-Sophia Antipolis, Laboratoire Physique de la Mati\`{e}re Condens\'{e}e, CNRS UMR 7336, 06100 Nice, France.}
\author{Gilles Montambaux}
\affiliation{Universit\'{e} Paris-Sud, Laboratoire de Physique des Solides, CNRS UMR 8502, 91405 Orsay Cedex, France.}
\author{Fabrice Mortessagne}
\email{fabrice.mortessagne@unice.fr}
\affiliation{Universit\'{e}  Nice-Sophia Antipolis, Laboratoire Physique de la Mati\`{e}re Condens\'{e}e, CNRS UMR 7336, 06100 Nice, France.}

\date{\today}

\begin{abstract}
By means of a microwave tight-binding analogue experiment of a graphene-like lattice, we observe a topological transition between a phase with a point-like band gap characteristic of massless Dirac fermions  and a gapped phase. By applying a controlled anisotropy on the structure, we investigate the transition  directly via  density of states measurements. The wave function associated with each eigenvalue is mapped and reveals new states at the Dirac point, localized on the armchair edges. We find that with increasing anisotropy, these new states are more and more localized at the edges.
\end{abstract}


\maketitle

{\em Introduction.}---Recently discovered condensed matter systems, such as graphene~\cite{CastroNeto2009} or topological insulators~\cite{Hasan2010}, constitute an ideal playground to investigate the physics of massless Dirac fermions until now restricted to high energy physics. Indeed, the relativistic spectrum emerges at the conical intersection points, the so-called Dirac points, in the low energy electronic dispersion relation. The large potential of technological applications~\footnote{For graphene, see a recent outlook in Nature \textbf{483}, S29--S74 (2012)} depends crucially on the properties of these quasi-particles. The manipulation of the Dirac points by external parameters -- from creation to annihilation -- have recently raised a strong interest both theoretically~\cite{Hasegawa2006,Zhu2007,*Wunsch2008,*Lee2009,Montambaux2009,*Montambaux2009a,Dietl2008} and experimentally~\cite{Tarruell2012,*Lim2012,Gomes2012}. By controlling the anisotropy of a honeycomb lattice, one can in principle  move the Dirac points up to a transition where they merge and annihilate each other. This is a \textit{topological transition} since both Dirac points are characterized by opposite topological numbers (opposite Berry phases) which annihilate at the transition. If the anisotropy is increased further, a band-gap opens in the dispersion relation. This transition from a gapless (Dirac) phase to a gapped phase corresponds to a Lifshitz phase transition from a semi-metallic to an insulating phase~\cite{Hasegawa2006,Zhu2007,*Wunsch2008,*Lee2009,Montambaux2009,*Montambaux2009a,Dietl2008}. The experimental investigation of this transition is difficult to perform with condensed matter systems, e.\,g.~graphene under uniaxial strain~\cite{Pereira2009} or quasi-two-dimensional organic conductors under pressure~\cite{Katayama2006,*Goerbig2008}. Artificial systems such as molecular graphene or ultracold atoms in optical lattices have been recently proposed to experimentally probe this topological transition~\cite{Tarruell2012,*Lim2012,Gomes2012}. While this effect has not been clearly observed with molecular graphene~\cite{Gomes2012}, Ref.~\cite{Tarruell2012,*Lim2012} reports the observation, using momentum-resolved interband transitions, of Dirac points merging with a Fermi gas in an anisotropic honeycomb lattice without six-fold symmetry.

Alternatively, discrete photonic systems (e.\,g.~photonic lattices or microwave cavities), with graphene-like structures, have been employed as condensed-matter analogues~\cite{Peleg2007,Bittner2010,*Zandbergen2010,*Bittner2012,Kuhl2010,Barkhofen2012a}. In microwave experiments, the tight-binding form of graphene's Hamiltonian~\cite{Wallace1947,*Reich2002} can be established by using a honeycomb lattice of evanescently coupled dielectric resonators~\cite{Kuhl2010,Barkhofen2012a}. Dirac points and signature of the linear dispersion relation have been observed~\cite{Bittner2010,*Zandbergen2010,*Bittner2012,Kuhl2010,Barkhofen2012a}.

In this Letter, we take advantage of the high versatility of our tight-binding microwave set-up to explore the topological phase transition. The anisotropy  of the honeycomb lattice is controlled through the nearest-neighbor hopping parameters. We directly measure the density of states (DOS) and the wave function associated with each eigenvalue. We observe a transition between the Dirac phase and the gapped phase exactly for the expected value of the anisotropy parameter defined in a tight-binding description. This is the first direct DOS measurement of this topological transition. Moreover the experimental set-up allows a very fine manipulation of the edges and a control of edge states which is not yet possible in condensed matter systems. We show that the anisotropy generates new states, at the Dirac frequency $\nu_D$, localized along specific armchair edges. With increasing anisotropy, their extension into the bulk decreases.
\medskip

{\em Experimental microwave graphene analogue.}---For our experiments, we use a set of 222 identical coupled dielectric cylindrical resonators (5 mm height, 8 mm diameter and a refractive index of 6), hereafter called \textit{discs}, placed in between two metallic plates. We establish a two-dimensional tight-binding regime where the electromagnetic field is mostly confined within the discs and spreads out evanescently. The experimental details are described in~\cite{Kuhl2010,Barkhofen2012a}. Via a movable loop-antenna, the reflected signal $S_{11}$ is measured over a given frequency range at the disc positions $\mathbf{r}$ using a vectorial network analyzer. The bare frequency $\nu_0$ of an isolated disc ($\sim 6.65$\,GHz in the set-up considered here) corresponds to the on-site energy appearing in the tight-binding Hamiltonian. The coupling parameter $t_{\mathrm{exp}}$ between two adjacent discs depends on the disc separation $d$ and can be described, as expected for a two-dimensional system, by a modified Bessel function $K_0$ (see Table.~\ref{ratio}).
Thus, by changing the distance between discs, one can change, in a controllable manner, the inter-site couplings. For a lattice composed of many discs, the set-up allows for a direct access to the \textit{local} density of states (LDOS) through the measured quantity $S_{11}$~\cite{Kuhl2010,Barkhofen2012a}. The DOS is obtained by averaging over all positions $\mathbf{r}$. Moreover, our set-up allows to visualize the wave function associated with each eigenfrequency~\cite{Bellec2012}.
\begin{figure}[t]
\includegraphics[scale=1]{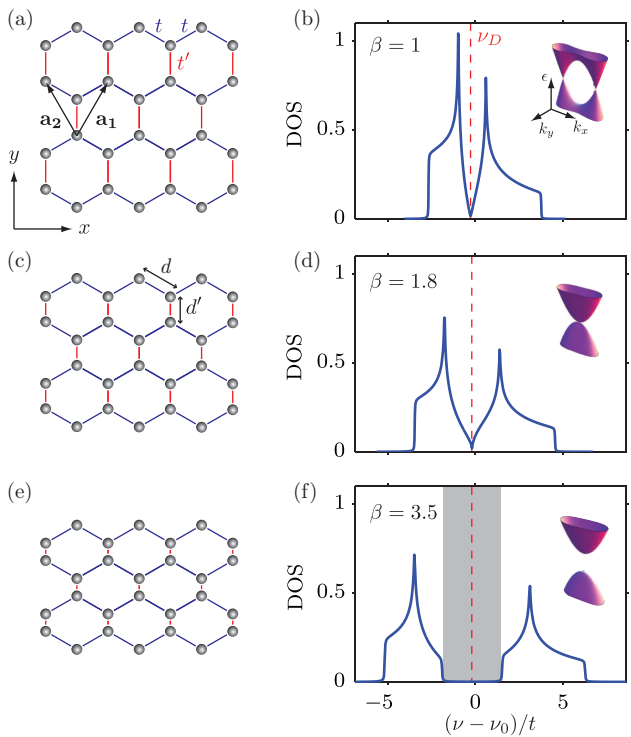}
\caption{\label{Sketch}(Color online)
Left panels: Schematics of the hexagonal lattices. $\mathbf{a_1}$ and $\mathbf{a_2}$ define the primitive cell of the Bravais lattice. $t$ (blue intersite links) and $t'$ (red ones, along $y$ axis) correspond to the nearest-neighbor couplings. $d$ and $d'$ are the corresponding distances. An anisotropy is introduced along the $y$ axis: $t'$ increases when $d'$ decreases. Right panels: Calculated density of states (DOS) for an infinite system with first, second and third nearest-neighbor couplings given in the text. The red dashed line is the Dirac frequency $\nu_D =\nu_0 + 3 t_2$. The insets show the corresponding dispersion relation. (a)-(b) Regular honeycomb lattice, $\beta=1$. The two Dirac points are distinguishable. (c)-(d) Compressed lattice at the topological transition, $\beta=1.8$. The Dirac points merge. (e)-(f) $\beta=3.5$, a band gap opens (gray zone).}
\end{figure}
\begin{figure}
\includegraphics[scale=1]{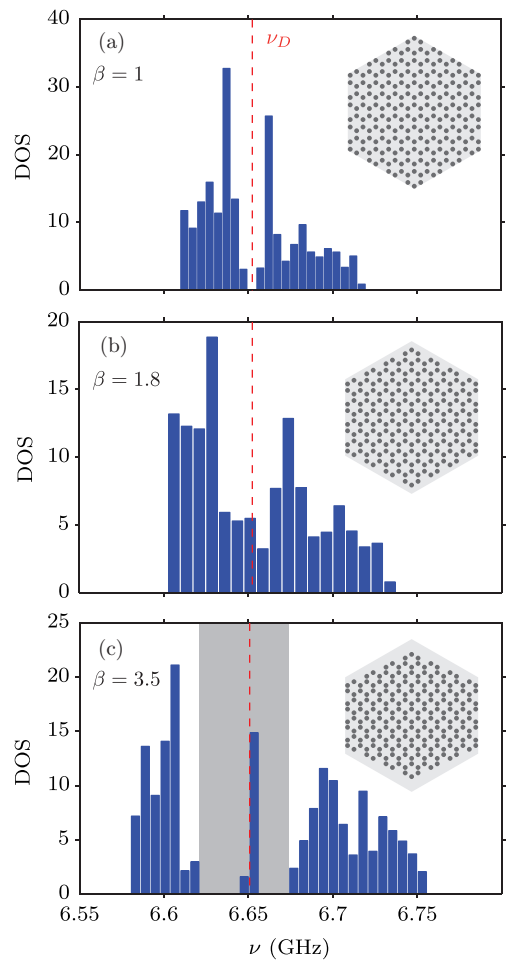}
\caption{\label{DOS1}(Color online)
Experimental DOS for (a) the regular graphene ($\beta=1$), (b) and (c) anisotropic structures with $\beta = 1.8$ and $\beta = 3.5$ respectively. The dashed red lines indicate  the Dirac frequency $\nu_D$. The gray zone shows the band gap with edge states at $\nu_D$.}
\end{figure}
\begin{table}
\caption{\label{ratio}
Experimental anisotropy parameter $\beta = t'_\mathrm{exp}/t_\mathrm{exp}$. For the regular lattice (i.\,e. $\beta = 1$), $d$ = 15 mm and $t_{\mathrm{exp}}$ = 0.016\,GHz. The intersite distances $d'$ and the corresponding nearest-neighbor coupling values $\beta t_\mathrm{exp}$ are also indicated.}
\begin{ruledtabular}
\begin{tabular}{lll}
$d'$ 	& $\beta t_\mathrm{exp}$ \footnotemark[1]	& $\beta$ 	\\
mm		& GHz										&  			\\
\hline
15.0						& 0.015				& \textbf{1}   	\\
13.9						& 0.020				& \textbf{1.3}  \\
13.0						& 0.027 			& \textbf{1.8}  \\
12.6						& 0.030				& \textbf{2}  	\\
12.0						& 0.037 			& \textbf{2.5} 	\\
11.0			 			& 0.053 			& \textbf{3.5} 	\\
\end{tabular}
\end{ruledtabular}
\footnotetext{Obtained according to the Bessel fit $t_{\mathrm{exp}}(d) = \alpha  \left| K_0 ( \gamma  d/2) \right|^2$ $+ \delta$. $\alpha = 1.9545$ GHz, $\gamma = 0.3167$ mm$^{-1}$ and $\delta = 0.0049$ GHz have been experimentally determined~\cite{Bellec2012}.}
\end{table}
\medskip

{\em Tight-binding description.}---In order to interpret quantitatively our experimental results, we present a tight-binding model on an infinite hexagonal lattice with first, second and third nearest-neighbor couplings. The on-site ``energy" is given by the bare frequency $\nu_0$. The hopping amplitudes (in units of frequency) between nearest-neighbor sites are called $t'$ along the $y$-axis and $t$ along the other directions (respectively red and blue links on Fig.~\ref{Sketch}(a). The next (next) nearest-neighbor couplings are respectively denoted as $t_2$ and $t_3$. The anisotropy of the lattice can be simply taken into account via the first nearest-neighbor coupling (i.\,e. $t'$), higher-order terms are assumed to be unchanged (i. e. $t_2 = t'_2$ and $t_3 = t'_3$). In the basis of the two inequivalent sites forming the elementary cell, the Hamiltonian can be written in the Bloch representation:
\begin{equation}
\label{hamilton}
\mathcal{H}_{\mathrm{TB}} = \begin{pmatrix}
\nu_0  + f_2(\mathbf{k}) 			& f(\mathbf{k}) +f_3(\mathbf{k})\\
f^*(\mathbf{k}) +f_3^*(\mathbf{k}) 	& \nu_0 + f_2(\mathbf{k})
\end{pmatrix}
\end{equation}
where $f$ (resp. $f_2$ and $f_3$) is the first (resp. second and third) nearest-neighbor contribution: $f(\mathbf{k}) = -(t'+t e^{i \mathbf{k} \cdot \mathbf{a}_1}+t e^{i \mathbf{k} \cdot \mathbf{a}_2})$, respectively $f_2(\mathbf{k}) = - 2 t_2 [\cos{\mathbf{k} \cdot \mathbf{a}_1} + \cos{\mathbf{k} \cdot \mathbf{a}_2}+ \cos{\mathbf{k} \cdot (\mathbf{a}_1-\mathbf{a}_2)}]$ and $f_3(\mathbf{k}) = - t_3 [ e^{i \mathbf{k} \cdot (\mathbf{a}_1 + \mathbf{a}_2)}+ e^{i \mathbf{k} \cdot (\mathbf{a}_1 - \mathbf{a}_2)} + e^{i \mathbf{k} (\mathbf{a}_2 - \mathbf{a}_1)}]$. The frequency spectrum is given by $\nu(\mathbf{k}) =\nu_0 + f_2(\mathbf{k}) \pm |f(\mathbf{k}) + f_3(\mathbf{k})|$ and consists of two bands touching at the Dirac frequency $\nu_D = \nu_0 + 3 t_2$. The corresponding DOS are represented in Fig.~\ref{Sketch} for different values of the anisotropy parameter $\beta = t'/t$. For $\beta=1$, two Dirac cones exist in $k$-space at $\mathbf{k} = \pm \mathbf{K}_D= \pm (\mathbf{a}_1^* - \mathbf{a}_2^*)/3$ where $\mathbf{a}_i^*$ are reciprocal lattice vectors. For larger $\beta$, the Dirac points moves along $k_x$ axis. A topological transition occurs at $\beta_c = 2-3t_3/t$, where the two Dirac points coalesce [inset Fig.~\ref{Sketch}(d)]. With the values of the coupling parameters extracted from the experiments (see below), we find $\beta_c=1.8$. For $\beta$ larger than this critical value, the condition $\nu(\mathbf{K}_D) = \nu_D$ cannot be fulfilled:  a gap opens [inset Fig.~\ref{Sketch}(f)]. As a consequence, the density of states is strongly affected. The typical linear relation at the vicinity of the Dirac point (red dashed line) observed for $\beta=1$ [Fig.~\ref{Sketch}(b)] turns to a square root law at the transition [Fig.~\ref{Sketch}(d)]~\cite{Hasegawa2006,Montambaux2009,*Montambaux2009a}. Then a gap opens for $\beta>\beta_c$ [gray zone in Fig.~\ref{Sketch}(f)]. Note that next (next) nearest-neighbor couplings lead to (\textit{i}) a lower value of $\beta_c$ for the transition (1.8 instead of 2  expected when only the first nearest-neighbor coupling is taken into account \cite{Montambaux2009,*Montambaux2009a,Hasegawa2006}), (\textit{ii}) an asymmetry of the bands,  (\textit{iii}) a slight shift of $\nu_D$ towards lower frequency and (\textit{iv}) a displacement of the two logarithmic singularities.
\medskip

{\emph{Topological transition and gap opening.}---Experimentally, we first realize a regular graphene lattice with an hexagonal shape which has only armchair boundaries in order to eliminate edge states [see inset Fig.~\ref{DOS1}(a)]. We gradually add an anisotropy by increasing the coupling $t'_{\mathrm{exp}}$ along the $y$ axis -- i.\,e. by decreasing the separation $d'$ between the  discs along $y$ (insets in Fig.~\ref{DOS1}). As detailed in the Tab.~\ref{ratio}, the anisotropy parameter $\beta$ varies from 1 to 3.5. The DOS corresponding respectively to $\beta=1,1.8,3.5$ are depicted in Fig.~\ref{DOS1}. For the regular graphene, for which $d = d' =15\,$mm, the non-symmetric band structure and the point-like gap at the Dirac frequency $\nu_D$ are observed [Fig.~\ref{DOS1}(a)]. Here, due to the finite size of the system, the expected linear behavior of the DOS near the Dirac point appears only as a dip in the histogram. Moreover, the two logarithmic singularities have their signature in this finite sample (Fig.~\ref{DOS1}). By adjusting the calculated DOS shown in Fig.~\ref{Sketch}(b) to the experimental one \cite{Bellec2012}, the following values for the three tight-binding couplings parameters are obtained: $t=0.016\,$GHz, $t_2/t=-0.091$ and $t_3/t=0.071$. This value of $t$ is in agreement with the Bessel fit of the experimental coupling (see Tab.~\ref{ratio}).\\
Fig.~\ref{DOS1}(b) and (c) show the evolution of the DOS when the lattice is compressed.  For large values of $\beta$, a gap opens as expected but additional states appear around $\nu_D$. We will provide a description of these states, identified as edge states, in the last part of the Letter. For an inspection of the topological transition, we eliminate their contributions by averaging the signal over bulk sites only.
\begin{figure}[b]
\includegraphics[scale=1.1]{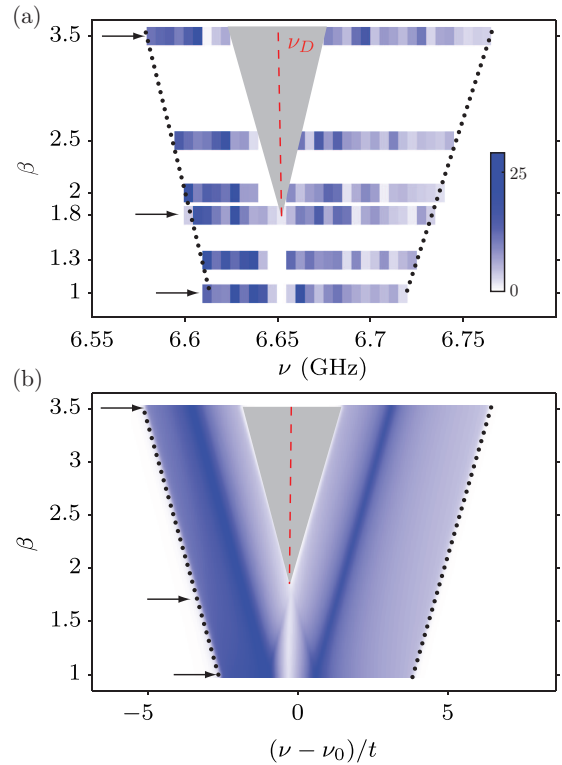}
\caption{\label{DOS2}(Color online)
(a) Experimental DOS for $\beta$ varying from 1 to 3.5. To remove edges states and highlight the band gap opening, the DOS have been averaged over bulk sites only. The arrows at $\beta = 1,1.8$ and 3.5 indicate the DOS plotted respectively in Fig.~\ref{DOS1}(a), (b) and (c), where the average was performed over all sites. (b) Calculated DOS for $\beta$ varying from 1 to 3.5 for an infinite system. The arrows at $\beta = 1,1.8$ and 3.5 indicate the DOS plotted in Fig.~\ref{Sketch}(b), (d) and (f) respectively. In (a) and (b), the gray zone shows the band gap opening, the dotted lines denote the linear dependence of the band edges and the red dashed line indicates $\nu_D$.}
\end{figure}
\begin{figure*}[t]
\includegraphics[scale=1]{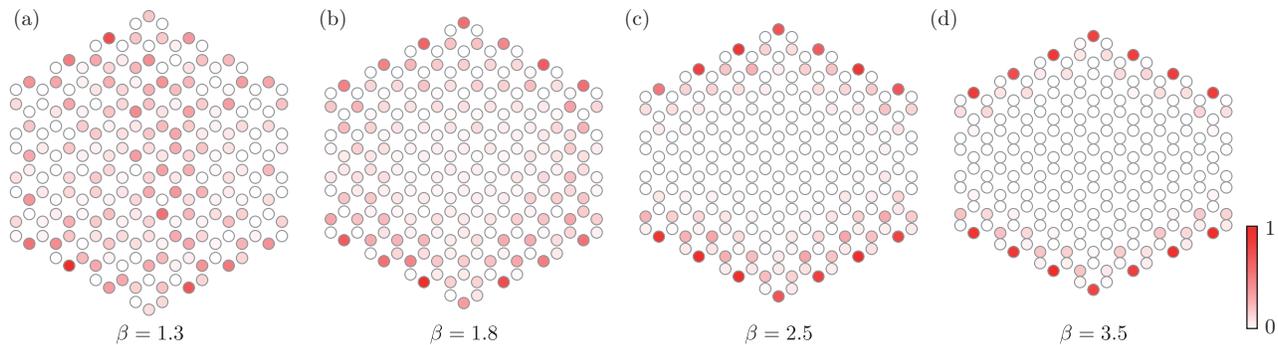}
\caption{\label{LDOS}(Color online)
Experimental wave function distributions at the Dirac frequency $\nu_D$ (red dashed line in Fig.~\ref{DOS2}) for $\beta$ ranging from 1.3 to 3.5.}
\end{figure*}
In Fig.~\ref{DOS2}(a), we plot the measured DOS obtained for six different values of $\beta$ using a white-blue colorscale. The dashed red line corresponds to the Dirac frequency and the gray zone is used here to improve the visualization of the band gap opening. These experimental results are successfully compared with the DOS calculated for the corresponding infinite systems [Fig.~\ref{DOS2}(b)]. We clearly distinguish the topological transition between a gapless phase and a gapped phase which occurs for $\beta_c = 1.8$, as expected from the theoretical expression $\beta_c=2-3t_3/t$ and the tight-binding parameters given above. Before the transition, the spectrum exhibits a point-like gap. At the transition, since the DOS close to the Dirac point is supposed to switch from a linear to a square root behavior~\cite{Hasegawa2006,Montambaux2009,*Montambaux2009a}, the number of states close to the Dirac point is expected to increase. Due to the finite size of the lattice, a small gap always exists, but a higher concentration of states at $\nu_D$ is clearly seen which reveals the topological phase transition. Finally, above the transition a gap opens. The gap and the bandwidths linearly increase with $\beta$ (respectively the gray zone and the dotted lines  in Fig. \ref{DOS2}).
\medskip

{\em Edge states.}---We now focus on the states which appear at the Dirac frequency $\nu_D$ for anisotropic lattices (see Fig.~\ref{DOS1}). The spatial distribution of these states is depicted in Fig.~\ref{LDOS} where it appears clearly that they are indeed edge states, which may appear {\it a priori} surprising  since armchair boundaries are not expected to support edge states~\cite{Nakada1996}. Their structure will be carefully studied in a future work, but several remarks are of immediate interest in this Letter. (\textit{i}) They appear only along the edges which are not parallel to the anisotropy axis. (\textit{ii}) Their localization along the edge increases when $\beta$ increases. (\textit{iii}) Their existence is not related to the topological transition observed in Fig.~\ref{DOS2}: they appear as soon as $\beta >1$. These features are in agreement with the prediction for the existence of armchair edge states in deformed structures~\cite{Delplace2011}. Moreover, we find that (\textit{i}) the intensity on one triangular sublattice stays zero, (\textit{ii}) the intensity on the other sublattice decreases roughly as $(1/ \beta)^{2 r}$ where $r$ is the distance to the edge in units of the lattice parameter. A more extensive investigation of these states, as well as of the states along zig-zag and  bearded edges in anisotropic structures is} in progress.
\medskip

{\em Conclusion.}---To conclude, we have exploited the high flexibility of a microwave two-dimensional  analogue of strained graphene to experimentally demonstrate the existence of a topological transition between a gapless phase and a gapped phase when the anisotropy increases. We have investigated the spectral properties of this structure by directly measuring the density of electromagnetic modes. An appropriate description of this system requires to take into account not only first but also next (next) nearest-neighbor couplings. The wave functions associated with the states appearing at the Dirac frequency are easily measurable with our set-up and they reveal the presence of edge states along the armchair edges. They become more and more localized along the edges as the anisotropy is increased. Finally, the versatility of the set-up allows to mimic many of the spectacular properties of graphene and  to reveal many other physical phenomena such as the influence of pseudo-magnetic field and associated Landau levels induced by a specific strain in graphene-like lattices~\cite{Guinea2009}.


%

\end{document}